\documentclass[12pt]{article}
\usepackage{amsmath}
\usepackage{latexsym}
\usepackage{amssymb}
\usepackage{fullpage}
 \usepackage{colortbl}
\usepackage{epsfig}
\usepackage{graphics}
\usepackage{fancyhdr}

\newcommand{\qed}{\hfill \ensuremath{\Box}}
\def\dimo{\noindent\mbox{\sc proof: }}
\def\eproof{\rm\hspace*{\fill}$\Box$\vspace{10pt}}
\newtheorem{defin}{\bf Definition}[section]
\newtheorem{theo}[defin]{Theorem}
\newtheorem{lem}[defin]{Lemma}
\newtheorem{corol}[defin]{Corollary}
\newtheorem{prop}[defin]{Proposition}

\newtheorem{oss}[defin]{Remark}
\newtheorem{ex}[defin]{Example}
\newtheorem{comment}[defin]{Comment}
\def\cF{{\cal F}}
\def\cW{{\cal W}}
\def\cG{{\cal G}}
\def\cP{\mbox{\boldmath ${\cal P}$}}

\def\double{\langle${\boldmath$\Pi$}, {\boldmath$\cF$}$\rangle}

\def\Pa{\{\phi=a\}}
\def\Pb{\{\phi=b\}}

\title{The structure of two-valued 
strategy-proof \\ social choice  functions with  indifference
}
\author{Achille Basile\thanks{Corresponding author.}\\
Dipartimento di Scienze Economiche e Statistiche \\
 Universit\`a Federico II, 
80126 Napoli, Italy\\
E-mail: basile@unina.it\\
\\
Surekha Rao \\
School of Business and Economics\\
Indiana University Northwest,
Gary, IN 46408\\
E-mail: skrao@iun.edu\\
and\\
K. P. S. Bhaskara Rao\\
Department of Computer Information Systems\\
Indiana 
University Northwest, Gary, IN 46408\\
E-mail: bkoppart@iun.edu}

\begin{document}
\maketitle

\thispagestyle{empty}

\begin{abstract}
 The paper investigates the structure of  coalitionally strategy-proof social choice functions -- CSP scfs, for short -- whose range is a subset of cardinality two of a given, arbitrarily large, set $A$ of alternatives. The study is conducted in the case where the voters/agents are allowed to express indifference among elements of $A$, and the domain  of the scfs consists of preference profiles  over a society $V$ of arbitrary cardinality. A representation formula for the two-valued CSP scfs is obtained that,  ultimately, provides the structure of such functions. Representation formulas are more specific than  characterization results since they de facto characterize, but  also furnish an operational way to construct all scfs under consideration.

\end{abstract}




 \newpage
 \setcounter{page}{1}
 
\section{Introduction}
\lhead{\sc \scriptsize The structure of two-valued ...}
\rhead{\sc  \scriptsize Introduction  }

 The purpose of this paper is to show a representation formula  for   strategy-proof social choice functions
 $$\phi: \cP\to A,$$
 under the following assumptions:
 \begin{itemize}
\item $A$ is a  set of alternatives having cardinality at least two;
 
\item $V$ is an arbitrary set of voters (i.e. the {\it society} $V$ is not necessarily finite);
 
\item $\cP$ is the set of all profiles $P=(P_v)_{v\in V}$ (universal domain hypothesis), consisting 
 of {\it weak orders} $P_v$  on  $A$ {\footnote{{\it Rational preference relations}, or just {\it preference relations}, according  to the books by Mas-Colell et al. \cite{MC} and by Aliprantis and Border \cite{AB} respectively.  
 So, throughout the sequel, a profile $P=(P_v)_{v\in V}$, consists of complete, transitive relations $P_v$  on a set $A$ of alternatives. }} - one for every member $v$ of the society;
 
\item the range of $\phi$ is of cardinality two.
 \end{itemize}

We adopt the  term  {\it two-valued social choice function} for a $\phi$ with range of cardinality two. If $A$ is itself of cardinality two and  $\phi$ is onto, we shall use the term {\it binary social choice function}. 
 We consider social choice functions  that are {\it coalitionally strategy-proof}, i.e. no group of agents has incentives to form a coalition that can  manipulate the social choice for their own advantage with false reporting (see Definition \ref{csp}). To be concise, sometimes we shall write scf to mean social choice function, and CSP to mean coalitionally strategy-proof.
 
  \bigskip
Representation formulas are more specific than  characterization results since they de facto characterize, but  also furnish an operational way to produce the desired scfs.
Therefore we are addressing a rather natural question and, more, we are considering this question even in the  simplest case of functions whose range is of cardinality two. Despite all this, until now not too many representation formulas can be found in the literature for tackling the case wherein voters may express indifference between alternatives.

When we assume that every voter can only express strict orderings, 
namely when we limit  the domain of $\phi$  by considering only strict profiles\footnote{If every $P_v$ is also antisymmetric, i.e. it is a {\it strict order} (one would say {\it strict preference} according to \cite{MC} or, reflexive {\it linear order} according to \cite{AB}), then we say that the profile P is strict.}, we have the following representation results (Theorem \ref{propo16} {\it (i)} and Theorem \ref{propo16} {\it (ii)}).

\begin{theo}\label{propo16}
The strategy-proof social choice functions defined on the set of all strict profiles, 
are all, and only, the following functions $P\mapsto \phi_{\cal F}(P)$.
\begin{itemize}

\item[(i)] Functions with  range $\{a, b\}\subseteq A$ of cardinality two:

\begin{equation*}
\phi_{\cal F}(P) = \left\{ 
\begin{array} {ll}
a,   & \mbox{ if } 
\{ v\in V: a\succ b, \mbox{ according to  } P_v \} 
\in \cF \\

 b,   & \mbox{ if }
 \{ v\in V: a\succ b, \mbox{ according to  } P_v\}  
 \notin \cF\,\,\\
\end{array}
\right.
\end{equation*}

where $\cF$ is a nonempty family of coalitions of voters which is closed under supersets.

\item[(ii)] Functions with finite range $A^*\subseteq A$ of cardinality three or more:
 
$$\phi_{\cal F} (P)=\mbox{ the unique alternative $a\in A^\star$ such that $D(a,P)\in {\cal F}$},$$
where ${\cal F}$ is an ultrafilter of coalitions, and, for  $x\in A^\star$, we denote by $D(x,P)$ the set   $\{v\in V: v \mbox{ has $x$ as his top choice  among the alternatives in {\it $A^\star$}, according to } P_v \}.
$
\end{itemize}

\end{theo}

The  formulation of $(i)$ and $(ii)$ can be found in Rao et al. \cite [Proposition 4.1,  and Theorem 3.5, respectively]{KPS}. In the case of a finite society $V$  and with  $A=\{a, b\}$, $(i)$ is due to Larsson and Svensson \cite[Theorem 2]{LS}\footnote{The Authors describe the rule $\phi_\cF$  as {\it voting by committees} (cfr. Peleg \cite{Peleg}, Barber\`a et al. \cite{BSZ}).}, whereas $(ii)$  combines results from Ishikava and Nakamura \cite {IN}, Mihara \cite {M1}, Pazner and Wesley \cite {PW}.

\bigskip

The scenario in which voters are allowed to express indifference among alternatives,  is much more complex. This is the setting of Barber\`a et al.  \cite{BBM}, where a strategy-proof characterization is offered in \cite[Theorem 1]{BBM} for two-valued social choice functions. To the best of our knowledge,  only recently Lahiri and Pramanik \cite[Theorem 1]{LP}  considered the problem of obtaining a representation formula. They 
gave a rather difficult formula for representing binary social choice functions in the framework of a finite society. For an arbitrary  society (i.e. finite or infinite) a simpler representation theorem for binary social functions has been obtained by Basile et al. \cite[Theorem 4.2]{KPS2}.  However, we do not know representation formulas when  indifference is permitted for cases such as functions with range of cardinality at least three and even  two-valued functions. In other words, we do not know results that generalize in a straightforward manner Theorem \ref{propo16} to the case of weak orderings. In the following we aim to give for the first time  a formula for  two-valued strategy-proof social choice functions. 
We will be considering the case of the range of cardinality three or more in a subsequent paper.

\bigskip
Facing two-valued functions rather than  merely binary ones, entails new serious difficulties that must be tackled.    This is so because the strategy-proofness is not sufficient to guarantee that the irrelevant alternatives, i.e. those not belonging to the range,  play no role.  In general, a strategy-proof social choice function defined on all weak profiles may not be independent of irrelevant alternatives. Take for example, $A = \{a, b, c\}$ and let one profile $P$ have every voter being indifferent between $a$ and $b$ but prefers them over $c$. Let a second profile $Q$ have every voter being indifferent between $a$ and $b$ but prefers $c$ over $a$ and $b$. Then $\phi(P)$ need not be equal to $\phi(Q)$ and such $\phi$ with range $\{ a, b\}$ can be easily constructed (see Example \ref{DIA}, {\it infra}).

In the binary setting a central role  is played by coalitions of indifferent agents. Here, due to the presence of further alternatives besides those in the range, we needed to give the analogous role to some profiles of preferences that we named partial indifference profiles. This is described in more detail below. A consistent extension to the present setting of the characterization of strategy-proofness by means of the concept of compatibility with dominance introduced in \cite{KPS2}, was hence necessary. Since our setting coincides with that of Barber\`a et al.  \cite{BBM},  we  also compare  their characterization result with ours.

In \cite{LS}, \cite{LP} and \cite{KPS2}, the representation theorems rely on the concept of nonempty superset closed families of coalitions (also named committees according to the Social Choice literature). Here a slight modification is necessary to treat the complexity of all possible situations: it is necessary to consider the possibility of empty families and also that a family contains the empty set, i.e.  it is the power set. Given that, the structure of two-valued strategy-proof scfs we have discovered in Definition \ref{defofpsi}, Theorems \ref{psiiscsp} and \ref{rappresentazione}, looks formally like  that of the binary particular case we achieved in  \cite[Theorem 4.2]{KPS2}. However, we wish to emphasize that it is by no means a straight extension of the previous representation, being the step from binary to two-valued functions quite considerable. 

Let us now illustrate the  structure of CSP scfs whose range is of cardinality two, say a subset $\{a, b\}$ of an arbitrary set $A$ of alternatives. For the purpose of giving a clearer intuition, we move along  the steps of the increasing difficulties of the scenarios we consider.

{\sc Strict profiles only.} Starting with aggregating strict profiles in a two-valued social choice, Theorem \ref{propo16} $(i)$ says that CSP scfs $\phi$ are identifiable with committees by means of the equation $\phi=\phi_\cF$. The fixed ultrafilters generated by the agents are among the possible committees, so that we are considering, as it is correct to do, also the so called dictatorial scfs. Since the latter do not exhaust all CSP scfs, if we want to describe all of them, we need to introduce functions such as $\phi_\cF$ that generalize dictatorial functions. 
This approach resembles Kirman and Sondermann \cite{KS}, wherein physical, visible, agents (i.e., fixed ultrafilters) are extended by adding invisible agents (i.e., arbitrary ultrafilters). For our purposes here, we need a further enlargement: We need to include committees.

{\sc Binary scfs.} Moving to binary scfs with indifference, a first point is that a single committee is not sufficient to give a CSP scf. However, a large class of CSP scfs is susceptible of a \lq\lq sequential\rq\rq\,\,description like the one that we are going to present.  We first expose the description in a simplified manner. 

\cite[Example 2.4]{KPS2}: 
Let $V$ be an at most countable set; enumerate its members as $v = 1, 2, ...$. Let  the scf be defined as follows: If the profile $P$ admits at least one agent for whom the two alternatives are not indifferent, then the social choice is dictated by the first (according to the natural ordering of $V$) non-indifferent agent; for the profile corresponding to the unanimous indifference, we assume the social choice is $x\in \{a, b\}$ (having fixed such an $x$). 


Mimicking the previous description,  for many  binary CSP scfs  $\phi$  one can find a well ordered collection $\cF_0, \cF_1, ..., \cF_\lambda, \dots$ of committees\footnote{In general $\lambda$ will describe an arbitrary well ordered set $\Lambda$} such that the social choice that aggregates a profile $P$ is that of the first (according to the ordering of $\Lambda$) committee which is not indifferent\footnote{For a committee $\cF$, the indifference between $a$ and $b$ means that the set $\{v\in V: a\quad {\underset {\sim}  \succ}_{P_v} b \}$ is in $\cF$ and, at the same time, the set $\{v\in V: b\quad {\underset {\sim}  \succ}_{P_v} a \}$ is in the  committee $\cF^\circ$ dual to $\cF$. Hence, when $\cF$ is not indifferent either we have that $\{v\in V: a\, {\underset {P_v}  \succ} \,b \}\in \cF$ (the committee choice is $a$) or we have that $\{v\in V: b\, {\underset {P_v}  \succ} \,a \}\in \cF^\circ$ (the committee choice is $b$).}, if there is one such committee; otherwise the choice will be the one that $\phi$ associates to the profile of unanimous indifference. This sequential structure is very simple and intuitive, and in \cite{KPS2} it is shown to capture many CSP scfs, but not all. Due to this difficulty (i.e. replacing committees with well ordered collections of committees is not sufficient) in \cite{KPS2} it has been proved that to obtain all binary CSP scfs, one can keep the  sequential structure described above as long as the role played by committees is played by  pairs $(I, \cF)$ consisting of a proper subset $I$ of $V$ where indifference is concentrated, and a committee on the coalition $V\setminus I$. This sounds not less intuitive than the sequential nature of a CSP scf. Indeed we are saying that once in a profile a set  (maybe empty) of agents is indifferent, the others will be requested to decide. For a pair $(I, \cF)$, that could be thought of as an \lq\lq extended committee\rq\rq,\,\, indifference or not must be adequately defined (cfr. Section 3 of \cite{KPS2}).

{\sc Two-valued scfs.} Since irrelevant alternatives count, it is clear that  the decisive role cannot be merely played by the mentioned pairs  $(I, \cF)$. It is exactly to tackle this difficulty that pairs $(\pi, \cF)$, i.e. a further enlargement of the notion of committee, comes into the scene: $\pi$ (partial indifference profile) is a profile not necessarily defined for all agents, its domain $I$ identifies  agents that are indifferent between $a$ and $b$, but also neutralizes  the role of the irrelevant alternatives by blocking the  preferences of the members of $I$ on them; as in the binary case, $\cF$ is a committee on $V\setminus I$. Giving, by means of  Definition \ref{classe}, a proper meaning to indifference or strict preference for a pair $(\pi, \cF)$,  our representation theorem reveals that the structure of all two-valued CSP scfs is the following: 
for every CSP scf $\phi$  one can find a well ordered collection $(\pi^0,\cF_0), (\pi^1,\cF_1), ..., (\pi^\lambda,\cF_\lambda), \dots$ of extended committees (which we name a  double collection, Definition \ref{doublecollection}) such that  the social choice that aggregates a profile $P$ is the choice of the first (according to the ordering of $\Lambda$)  extended committee which is not indifferent, if there is one; otherwise the choice will be the one that $\phi$ associates to a profile of unanimous indifference between $a$ and $b$.

Moreover we also point out  that the possibility of fixing a priori, freely, the parameter
$x$ of the representation\footnote{One can choose ex ante $x$ as the value of the scf on any profile of unanimous indifference between the alternatives in the range.} has to be considered jointly with the fact that the other parameter - the double collection $\double$  - will  change coherently.

\bigskip
The paper is organized as follows. In Section 2 we introduce the dominance relation between profiles and the concept of compatibility of a social choice function with the dominance. The characterization of strategy-proofness by means of  compatibility is in Theorem \ref{final2} and the Section ends with a comparison of compatibility with the notions of essentially ab-based and ab-monotonic due to Barber\`a et al \cite{BBM}. Section 3 presents some notions which are preparatory to the definition of index of a profile, and the index itself. This is a tool for the definition in Section 4 of some special scfs of sequential form, that we call of type $\psi$. Theorem \ref{psiiscsp} shows that such functions are strategy-proof, whereas to the fact that there are no other two-valued strategy-proof scfs is devoted Section 5. The general representation result is Theorem \ref{rappresentazione}, whose proof is given in the final Appendix. 

\newpage
\section{Strategy-proofness}
\lhead{\sc \scriptsize The structure of two-valued ...}
\rhead{\sc\scriptsize Strategy-proofness}

Let $\cW(A)$ be the set of weak orderings over the set $A$ of alternatives. Functions $\pi$ from subsets of $V$ to $\cW(A)$ are named partial profiles of preferences. A partial profile is therefore $\pi=(\pi_v)_{v\in I}$ where $I$ is a subset of $V$ and $\pi_v\in \cW(A)$ for every $v \in I$.
In case  $I=V$, we speak of a (total) profile of preferences, omitting \lq\lq total\rq\rq. For the class of all (total) profiles we use the notation $\cP$. For the case in which the domain  is the empty set, we reserve the name of empty profile. 
For a profile $P=(P_v)_{v\in V}$ we shall also use the notation  $P=[P_I, P_{I^c}]$   if $I$ is a subset of $V$ and $P_I$, $P_{I^c}$ are the obvious restrictions $P_I=(P_v)_{v\in I}$, $P_{I^c}=(P_v)_{v\notin V}$ of $P$\footnote{We denote by the superscript $^c$ the complement of a set.}. Extending this notation to arbitrary partitions of $V$ or to partial profiles is straightforward.

With reference to   an ordering  $W$ on the set $A$ of alternatives,  as usual, the notation $ x\quad {\underset {\sim}  \succ}_{W} y$  stands for $(x,y)\in W$, the notation $ x\, \underset W{\succ} \,y$ stands for $\big[(x,y)\in W$ and $(y,x)\notin W\big]$, and the notation $ x\, \underset W{\sim} \,y$ stands for $\big[(x,y)\in W$ and $(y,x)\in W\big]$.

\begin{defin}\label{partial}
Let $a$ and $b$ be two distinct alternatives. A {\bf partial $\{a, b\}$-indifference profile} is a partial profile $\pi$ such that for every voter $v\in$ {\rm dom }$(\pi)$  (the domain  of $\pi$) one has \quad
$a \underset{\pi_v}\sim b.$
\end{defin}
Trivially, the empty profile is a partial $\{a, b\}$-indifference profile. The class of all  partial $\{a, b\}$-indifference profiles will be denoted by $\cP(a\sim b)$. 
It contains profiles of unanimous indifference between $a$ and $b$  also, namely profiles $P=(P_v)_{v\in V}$ such that $a \underset{P_v}\sim b, \,\, \forall v\in V.$
Notice that if $A$ consists of two alternatives $a, b$ only, then $\cP(a\sim b)$ can be identified, via the mapping $\pi\mapsto {\rm dom}(\pi)$, with the power set of $V$.

\bigskip
Throughout the sequel,  $\phi: \cP\to A$ stands for a scf. Dealing with societies of any cardinality, the notion of strategy-proofness we adopt is based on coalitions of individuals rather than on a single individual. As usual,  coalition is  synonymous to  nonempty subset of $V$. 

\begin{defin}\label{manipulation}
Let $\phi$ be scf.
 We say that a coalition $D$ can manipulate a profile $P$   under $\phi$  if there is another profile $Q$ such that 

\begin{itemize}
\item every voter $v$ in $D^c$ has the same preference order in both $P$ and $Q$, \, i.e. $P_v=Q_v$; 

\item every voter $v$ in $D$ prefers $\phi(Q)$ to $\phi(P)$ according to $P_v$, \, i.e. $\phi(Q)\underset{P_v}\succ \, \phi(P)$.
\end{itemize}
When the two conditions above are verified we say that  the coalition $D$ manipulates the proflile $P$ by presenting (or reporting) the profile $Q$.
\end{defin}

A desirable property of a scf  is that no group of agents has incentives to form a coalition that, with false reporting, can manipulate the social outcome for their own advantage. This is at the basis of  the following  definition.

\begin{defin}\label{csp}
 We say that a scf $\phi$ is  {\bf  coalitionally strategy-proof} (CSP, for short) if no coalition of voters can manipulate any profile under $\phi$. 
\end{defin}

If in the previous definitions we replace arbitrary coalitions with singletons, the notion we identify is that of
{\it individual strategy-proofness}.

It is obvious that every CSP social choice function is individually strategy-proof. On the other hand it is well known that the two concepts of strategy-proofness, individual and coalitional, coincide for the case of $V$ being finite, whereas they  do not coincide if $V$ is infinite. 

\begin{ex}\label{DIA}
{\rm
The purpose of this example is to clarify that strategy-proofness does not imply independence of irrelevant alternatives.  
Let $V, A$ be consisting of two agents $\{v_1, v_2\}$ and three alternatives $\{a, b, c\}$.
A strategy-proof scf $\phi$ with range $\{a, b\}$  can be defined as follows.  If for the profile $P$ either one of the following 

$$ b \underset{P_{v_1}} \succ a; \qquad c \underset{P_{v_1}} \succ a \underset{P_{v_1}}\sim b;  \qquad  a \underset{P_{v_1}}\sim b \underset{P_{v_1}} \succ c  \,\,\&\,\,  b \underset{P_{v_2}} \succ a;$$

holds true, then we define $\phi(P)=b$. In all the other cases we define $\phi(P)=a$.

Trivially, if we take the profiles $P$ and $Q$ as below

$$\begin{tabular}{llll}
&  $\ \ \ \ \ v_{1}$ &  $\ \ \ \ \ v_{2}$ &  \\ \cline{2-3}
Profile \  $P$ & \multicolumn{1}{|l}{$a\sim b\succ c$} & \multicolumn{1}{|l}{%
$a\sim b\succ c$} & \multicolumn{1}{|l}{$\overset\phi\longrightarrow a$} \\ \cline{2-3}
&  &  &  \\ \cline{2-3}
Profile \  $Q$ & \multicolumn{1}{|l}{$c\succ a\sim b$} & \multicolumn{1}{|l}{%
$c\succ a\sim b$} & \multicolumn{1}{|l}{$\overset\phi\longrightarrow b$} \\ \cline{2-3}
\end{tabular}
$$
we have that $\phi(P)=a$ and $\phi(Q)=b$.  We have a similar outcome  if we take $P$ and $Q$ such that
$$
\begin{tabular}{llll}
&  $\ \ \ \ \ v_{1}$ &  $\ \ \ \ \ v_{2}$ &  \\ \cline{2-3}
Profile \  $P$ & \multicolumn{1}{|l}{$a\sim b\succ c$} & \multicolumn{1}{|l}{%
$a\succ b\succ c$} & \multicolumn{1}{|l}{$\overset\phi\longrightarrow a$} \\ \cline{2-3}
&  &  &  \\ \cline{2-3}
Profile \  $Q$ & \multicolumn{1}{|l}{$c\succ a\sim b$} & \multicolumn{1}{|l}{%
$a\succ b\succ c$} & \multicolumn{1}{|l}{$\overset\phi\longrightarrow b$} \\ \cline{2-3}
\end{tabular}
$$
This is so even if the function $\phi$ is strategy-proof and the profiles $P$ and $Q$ are identical on $\{a, b\}$. The fact that $\phi$ cannot be manipulated can be easily shown. Suppose, on the contrary, that a voter $v$  can manipulate a profile $P$ reporting a profile  
$Q$, we have: $\phi(Q)\underset{P_v}\succ \phi(P)$ and $P_{-v}=Q_{-v}$. Without loss of generality we can suppose that $\phi(Q)=b$. Hence we have $b=\phi(Q)\underset{P_v}\succ \phi(P)=a$. The voter $v$ cannot be $v_1$ otherwise $\phi(P)$ is $b$ by definition of $\phi$. Hence we have that $v=v_2$ and therefore that $b=\phi(Q)\underset{P_{v_2}}\succ \phi(P)=a$ and $P_{{v_1}}=Q_{{v_1}}$. By the definition of $\phi$, the fact that 
$\phi(Q)=b$, $P_{{v_1}}=Q_{{v_1}}$, and $\phi(P)=a$, we have that necessarily $a \underset{P_{v_1}}\sim b \underset{P_{v_1}} \succ c$ but this together with $b\underset{P_{v_2}}\succ a$ would give that $\phi$ on $P$ takes value $b$, a contradiction.\eproof
}\end{ex}

\begin{oss}\label{Pareto}
{\sl It is immediate that a scf $\phi$ which is CSP, enjoys  
 the following property:
$$
\{y\in \mbox{Range}(\phi): y\underset{P_v}\succ \phi(P), \forall v\in V \}=\O, \mbox{ for every profile }P\in\mbox{ {\rm dom}}(\phi)
$$
which is the well known weak {\bf Pareto optimality}.}
\end{oss}

\bigskip
Since our purpose is to find a representation formula for CSP scfs that have range of cardinality two, 
we adopt the following approach: {\bf we fix a set $\{a, b\}\subseteq A$ consisting of two distinct alternatives and produce a formula for CSP social choice functions with range  $\{a, b\}$}.
With this in mind,   we introduce the following notations. Let  $a$ and $b$ be distinct alternatives and $P$ be a profile, then we set:
  
  \begin{itemize}
\item [--]  the indifferece set of voters: \quad $I(P)=\{v\in V: a \underset{P_v}\sim b\},$  
  \,
\item [--] the set of voters that prefer $a$ to $b$:  \quad $D(a, P)=\{v\in V: a \underset{P_v}\succ b\},$
  
 \item [--] the set of voters that prefer $b$ to $a$: \quad  $D(b, P)=\{v\in V: b \underset{P_v}\succ a\}.$

 \end{itemize}
 
 Also useful is the following notion of  $\{a,b\}$-equivalence set of two profiles $P$ and $Q$.
  
  \begin{defin}
 Let $P,Q$ be two profiles.
 The {\bf $\{a, b\}$-equivalence set of $P$ and $Q$}  consists of the voters $v\in V$ for which  either one of the following three statements hold true:
\begin{itemize}
\item $P_v=Q_v$;
\item  restricted to $\{a, b\}$,  the preferences $P_v$ and $Q_v$ are identical and coincide with $a\succ b$;
\item restricted to $\{a, b\}$,  the preferences $P_v$ and $Q_v$ are identical and coincide with $b\succ a$.
\end{itemize}
The $\{a, b\}$-equivalence set of $P$ and $Q$ will  denoted by  $E(P,Q)$.
  \end{defin}

\begin{comment}{\bf(about notation)}\label{abuse}
{\sl For the four sets we have introduced, a more appropriate notation should involve the set $\{a, b\}$ also:\,\,
 $I_{\{a,b\}}(P)$, $D_{\{a,b\}}(a, P)$, $D_{\{a,b\}}(b, P)$, $E_{\{a,b\}}(P,Q)$. The notational abuse  is for the sake of simplicity.
Throughout the paper the reference to  the two alternatives $a$ and $b$ will be unambiguous. \eproof
}\end{comment}

We now introduce the following dominance relation between profiles.

\begin{defin}\label{dominance}
Let ${\{a,b\}}$ be two distinct alternatives in $A$. Given two profiles $P$ and $Q$, we say that $P$ dominates $Q$ and write $P\overset{{\{a,b\}}}\sqsupset Q$ if either $V= E(P,Q)\cup \big[I(P)\cap D(a, Q)\big]$ 
or $V= E(P,Q)\cup \big[I(P)\cap D(b, Q)\big]$.
\end{defin}

{\bf Comment \ref{abuse} (continuation)}
{\sl It is for the already invoked sake of simplicity that we do not explicitly speak of ${\{a,b\}}$-dominance and, moreover,  we write $P\sqsupset Q$ instead of $P\overset{{\{a,b\}}}\sqsupset Q$.

\smallskip
When  in particular we have $V= E(P,Q)\cup \big[I(P)\cap D(a, Q)\big]$, we adopt the notation $P{\overset{\{a,b\}}{\underset a\sqsupset}} Q$ that, however, we simplify as $P\underset a\sqsupset Q$. In this case we shall say that the profile $P$ $b$-dominates the profile Q.

Analogously, for  $V= E(P,Q)\cup \big[I(P)\cap D(b, Q)\big]$ we write  $P{\overset{\{a,b\}}{\underset b\sqsupset}} Q$ that we simplify as
 $P\underset b\sqsupset Q$. We shall say in this case that the profile $P$ $b$-dominates the profile Q.
}\eproof

\bigskip

Clearly the unions in the definition of dominance are disjoint unions. Notice that the following are obviously equivalent:
\begin{itemize}
\item The simultaneous validity of the two dominances, $P\underset a\sqsupset Q$  and $P\underset b\sqsupset Q$,

\item The simultaneous validity of the two dominances, $P\underset a\sqsupset Q$  and $Q\underset a\sqsupset P$,

\item The simultaneous validity of the two dominances, $P\underset b\sqsupset Q$  and $Q\underset b\sqsupset P$,

\item $V=E(P,Q)$
\end{itemize}

 So in general
the dominance $\underset a\sqsupset$ (and $\underset b\sqsupset$) is not antisymmetric ($P\underset a\sqsupset Q$ and $Q\underset a\sqsupset P\not\Rightarrow P=Q$) although it is reflexive and transitive (not complete). 
%

\begin{defin}\label{funzioni compatibili}
Let $\phi$ be a two-valued scf, and let $\{a, b\}$ be its range.  
We say that $\phi$ is compatible with the dominance relation $\sqsupset$ when 

$$
P\underset {\phi(P)}\sqsupset Q\,\Rightarrow \,\, \phi(Q)=\phi(P). 
$$
\end{defin}
We shall now characterize coalitional strategy-proofness  in terms of dominance.

\begin{theo}\label{final2}

A scf $\phi$ with range $\{a, b\}$ is CSP if and only if it is compatible with  $\sqsupset$.
\end{theo}


\bigskip

\dimo
Suppose $\phi$ is compatible with $\sqsupset$ and assume that it is not CSP.
This means that  profiles $P, Q$ exist and a coalition $D$ such that  $P$ and $Q$ are identical on $D^c$ and for $v\in D$ we have $\phi(Q) \underset{P_v}\succ \phi(P)$.

Let us fix, without loss of generality, that $\phi(P)=a$ and $\phi(Q)=b$. Define the following partition of $D$:
$$D_1= D(b, Q)\cap D, \quad D_2=D(a, Q)\cap D, \quad D_3=I(Q)\cap D.$$
and a new profile $R=[R_{D^c}, R_{D_1}, R_{D_2}, R_{D_3}]$,   identical to $P$ (and $Q$)
on $D^c$,  to $P$ on $D_1$, to $Q$ on $D_3$, and arbitrary on $D_2$ as long as it guarantees $a \underset{R_v}\sim b$ for $v\in D_2$. The figure below describes the situation.


\bigskip
$$
\begin{tabular}{llllll}
\cline{1-5}
\multicolumn{1}{|l}{Profile  $P$} & \multicolumn{1}{|l}{$P$} & 
\multicolumn{1}{|l}{$b\succ a$} & \multicolumn{1}{|l}{$b\succ a$} & 
\multicolumn{1}{|l}{$b\succ a$} & \multicolumn{1}{|l}{$\overset\phi\longrightarrow a$}
\\ \cline{1-5}
&  &  &  &  &  \\ \cline{1-5}
\multicolumn{1}{|l}{Profile  $Q$} & \multicolumn{1}{|l}{$P$} & 
\multicolumn{1}{|l}{$b\succ a$} & \multicolumn{1}{|l}{$a\succ b$} & 
\multicolumn{1}{|l}{$a\sim b$} & \multicolumn{1}{|l}{$\overset\phi\longrightarrow b$} \\ 
\cline{1-5}
& \multicolumn{1}{|l}{} & \multicolumn{1}{|l}{} & \multicolumn{1}{|l}{} & 
\multicolumn{1}{|l}{} &  \\ 
& \multicolumn{1}{|l}{$D^{c}$} & \multicolumn{1}{|l}{ $\ D_{1}$} & 
\multicolumn{1}{|l}{ $\ D_{2}$} & \multicolumn{1}{|l}{ $\ D_{3}$} &  \\ 
& \multicolumn{1}{|l}{} & \multicolumn{1}{|l}{} & \multicolumn{1}{|l}{} & 
\multicolumn{1}{|l}{} &  \\ \cline{1-5}
\multicolumn{1}{|l}{Profile  $R$} & \multicolumn{1}{|l}{$P$} & 
\multicolumn{1}{|l}{ $\ P$} & \multicolumn{1}{|l}{$a\sim b$} & 
\multicolumn{1}{|l}{ $\ Q$} & \multicolumn{1}{|l}{} \\ \cline{1-5}
\end{tabular}
$$
\bigskip

It is straightforward to observe that 
$$
E(R,Q)= D^c\cup D_1 \cup D_3; \qquad E(R,P)= D^c\cup D_1. 
$$
This entails
$$
R\underset a\sqsupset Q\,\mbox{ and } R\underset b\sqsupset P.
$$
Because of the compatibility assumption, $\phi(R)$ must be $a$ and $b$ at the same time, this produces a contradiction.
Thus, if $\phi$ is compatible with $\sqsupset$ then it is CSP.

\bigskip

Now we prove the converse. Let us suppose that $\phi$ is CSP. We have to prove that
$$
P\underset {\phi(P)}\sqsupset Q\,\Rightarrow \,\, \phi(Q)=\phi(P). 
$$
Since $\phi(P)\in \{a, b\}$, we consider the case $\phi(P)=a$. The other case can be treated similarly. Let us consider the following partition $\{V_1, V_2, V_3, V_4\}$ of $V$:

\medskip
$V_1=\{v\in V: P_v=Q_v\}$, \quad $V_2=\{v\in V: P_v\neq Q_v; {P_v}_{|\{a, b\}}={Q_v}_{|\{a, b\}}=a\succ b\}$,

\medskip
$V_3=\{v\in V: P_v\neq Q_v; {P_v}_{|\{a, b\}}={Q_v}_{|\{a, b\}}=b\succ a\}$, \quad
$V_4= V\setminus E(P,Q),$ 

\medskip
since clearly $\{V_1, V_2, V_3\}$ is a partition of $E(P,Q)$.
The figure below illustrate the situation we have:

\medskip
$P=[P_{V_1}, P_{V_2}, P_{V_3},P_{V_4}]$;\quad
$Q=[P_{V_1}, Q_{V_2}, Q_{V_3},Q_{V_4}]$;\quad
$P_{V_2}\neq Q_{V_2}; \quad P_{V_3}\neq Q_{V_3}$;

\medskip
$\big[v\in V_2\Rightarrow  a\underset{P_v}\succ b \,\, \& \,\, a\underset{Q_v}\succ b\big]$;\quad 
$\big[v\in V_3\Rightarrow  b\underset{P_v}\succ a \,\, \& \,\, b\underset{Q_v}\succ a\big]$;\quad
$\big[v\in V_4\Rightarrow  a\underset{P_v}\sim b \,\, \& \,\, a\underset{Q_v}\succ b\big].$

\bigskip

$$
\begin{tabular}{llllll}
\cline{1-5}
\multicolumn{1}{|l}{Profile $P$} & \multicolumn{1}{|l}{$P$} & 
\multicolumn{1}{|l}{$a\succ b$} & \multicolumn{1}{|l}{$b\succ a$} & 
\multicolumn{1}{|l}{$a\sim b$} & \multicolumn{1}{|l}{$\overset\phi\longrightarrow a$} \\ 
\cline{1-5}\cline{3-4}
&  & \multicolumn{1}{|l}{$P\neq Q$} & \multicolumn{1}{|l}{$P\neq Q$} & 
\multicolumn{1}{|l}{} &  \\ \cline{1-5}\cline{3-4}
\multicolumn{1}{|l}{Profile $Q$} & \multicolumn{1}{|l}{$P$} & 
\multicolumn{1}{|l}{$a\succ b$} & \multicolumn{1}{|l}{$b\succ a$} & 
\multicolumn{1}{|l}{$a\succ b$} & \multicolumn{1}{|l}{} \\ \cline{1-5}
& \multicolumn{1}{|l}{} & \multicolumn{1}{|l}{} & \multicolumn{1}{|l}{} & 
\multicolumn{1}{|l}{} &  \\ 
& \multicolumn{1}{|l}{ $\ V_{1}$} & \multicolumn{1}{|l}{ $\ V_{2}$} & 
\multicolumn{1}{|l}{ $\ V_{3}$} & \multicolumn{1}{|l}{ $\ V_{4}$} &  \\ 
&  &  &  &  & 
\end{tabular}
$$

\bigskip

We assume $V_1$ is a proper subset of $V$, otherwise the assertion we need to prove, i.e. $\phi(Q)=a$,  would be trivial since $P$ and $Q$ would coincide. One of the sets
$\{V_2, V_3, V_4\}$ is nonempty. Without loss of generality we assume $V_2$ is nonempty, and, in  every one  of the four possible cases determined by the nonemptyness of $V_3, V_4$, we shall prove the assertion.

\medskip
Case $V_3=V_4=\O.$

We have $P=[P_{V_1}, P_{V_2}]$ and $Q=[P_{V_1}, Q_{V_2}]$, therefore if $\phi(Q)=b$, the coalition $V_2$ manipulates $Q$ presenting $P$.

\medskip
Case $V_3=\O; V_4\neq\O.$

We have $P=[P_{V_1}, P_{V_2}, P_{V_4}]$ and $Q=[P_{V_1}, Q_{V_2}, Q_{V_4}]$. Define the profile $S=[P_{V_1}, Q_{V_2}, P_{V_4}]$ over which the value of $\phi$ must be $a$ otherwise $V_2$ presents  $P$ and manipulates $S$. At this point also $\phi(Q)=a$ otherwise the coalition $V_4$ can manipulate $Q$ presenting $S$.

\medskip
Case $V_3\neq \O; V_4=\O.$

We have $P=[P_{V_1}, P_{V_2}, P_{V_3}]$ and $Q=[P_{V_1}, Q_{V_2}, Q_{V_3}]$. 
Define the profile $S=[P_{V_1}, Q_{V_2}, P_{V_3}]$ over which the value of $\phi$ must be $a$ otherwise $V_2$ presents  $P$ and manipulates $S$. At this point also $\phi(Q)=a$ otherwise the coalition $V_3$ can manipulate $S$ presenting $Q$.

\medskip
Case $V_3\neq \O; V_4\neq\O.$

In this last case we see that $\phi([P_{V_1}, Q_{V_2}, P_{V_3},P_{V_4} ])=a$, otherwise  the coalition $V_2$ manipulates $[P_{V_1}, Q_{V_2}, P_{V_3},P_{V_4} ]$ presenting $P$.
Hence, necessarily
$\phi([P_{V_1}, Q_{V_2}, P_{V_3},Q_{V_4} ])=a$, otherwise $V_4$ manipulates $[P_{V_1}, Q_{V_2}, P_{V_3},Q_{V_4} ]$ presenting $[P_{V_1}, Q_{V_2}, P_{V_3},P_{V_4} ]$.
Finally we get $\phi(Q)=a$, otherwise if $\phi(Q)=b$, $V_3$ manipulates $[P_{V_1}, Q_{V_2}, P_{V_3},Q_{V_4} ]$ presenting $Q$.
\eproof 

\begin{corol}
Let $\phi$ be a CSP scf with range $\{a, b\}$. Then
$$
V=E(P,Q) \Longrightarrow  \phi(P)=\phi(Q).
$$
\end{corol}
 
 \begin{oss} {\sl
 
In  Barber\`a et al \cite{BBM}, where the society $V$ is finite, the notion of coalitional strategy-proofness is called {\it weak group strategy-proofness} and it is characterized as follows.

\cite [Theorem 1]{BBM}: \quad 
{\it if the range of $\phi$ is $\{a, b\}$, then $\phi$ is CSP if and only if  it is {\bf essentially ab-based} and {\bf essentially ab-monotonic}}.

The purpose of this Remark is to show  directly that a function $\phi$ with range $\{a, b\}$ has both the above properties if and only if it is compatible with the $\{a, b\}$-dominance relation.

 \bigskip
 We remind that according to \cite{BBM}, 
\begin{itemize}
\item $\phi$ is essentially ab-based, if the following implication holds true:

 $$
  \left. 
\begin{array} {ll}
P, Q\in\cP; \,\, v\in I(P)\Rightarrow P_v=Q_v; \\
D(a,P)= D(a,Q); \,\,
D(b, P) = D(b, Q)  \\
\end{array}
\right]\Rightarrow  \,\,\phi(P)=\phi (Q);
 $$

\item $\phi$ is essentially ab-monotonic if the following implications (1) and (2) hold true:
 
$$
  (1)\left[ 
\begin{array} {ll}
P, Q\in\cP; \, v\in I(P)\cap I(Q)\Rightarrow P_v=Q_v; \, \phi(P)=a; \\
D(a,Q)\supseteq D(a,P); 
D(b, P)\supseteq D(b, Q)  \mbox{(with at least  }\\  \mbox{ one strict inclusion)}\\
\end{array}
\right]\Rightarrow  \phi(Q)=a,
 $$

$$
 (2) \left[ 
\begin{array} {ll}
P, Q\in\cP; \, v\in I(P)\cap I(Q)\Rightarrow P_v=Q_v; \, \phi(P)=b; \\
D(b,Q)\supseteq D(b,P); 
D(a, P)\supseteq D(a, Q)  \mbox{ (with at least}\\ \mbox{ one strict inclusion)}\\
\end{array}
\right]\Rightarrow  \phi(Q)=b.
 $$
\end{itemize}
Now we observe that the simultaneous validity of essential ab-based and ab-monotonic conditions can be  restated as the validity of both implications (1) and (2) above, without requiring that at least one of the inclusions involved is strict (see the implications $(1')$ and $(2')$ below). We shall in fact show that any one of the two modified implications suffices.

To see this, let us write $B_i(P,Q)$, for a pair of profiles $(P,Q)$,  if the three conditions (distinct for $i=1$ and $i=2$) below hods true:

$$\begin{tabular}{cc||c}
& $B_{1}(P,Q)$ & $B_{2}(P,Q)$ \\ \cline{2-3}
 & \multicolumn{2}{c}{$v\in I(P)\cap I(Q)\Rightarrow P_{v}=Q_{v}
$} \\ 
 & $D(a,Q)\supseteq D(a,P)$ & $D(a,P)\supseteq D(a,Q)$ \\ 
 & $D(b,P)\supseteq D(b,Q)$ & $D(b,Q)\supseteq D(b,P)$%
\end{tabular}$$

Given that, and observing that $B_2(P,Q)$ is the same as $B_1(Q,P)$, we can  conclude in a straightforward manner that

\begin{prop}\label{essential}
$\phi$ is essentially ab-based and essentially ab-monotonic if and only if 
the following implications $(1')$ and $(2')$ hold true simultaneously:
$$
  (1') \qquad \big[B_1 (P,Q) ,\, \phi(P)=a\big]\Rightarrow  \phi(Q)=a,
 $$

$$
   (2') \qquad \big[B_2 (P,Q) ,\, \phi(P)=b\big]\Rightarrow  \phi(Q)=b.
 $$



Moreover, the simultaneous validity of both $(1')$ and $(2')$  is equivalent to the validity of either $(1')$ or $(2')$.
\end{prop}

On the basis of the previous Proposition, we can proceed to the  direct comparison between compatibility with dominance and essential ab-based and ab-monotonic conditions.

\begin{prop}
Let $\phi$ be a scf with range $\{a, b\}$\footnote{Remind that V is not necessarily  finite.}. Then, $\phi$ is essentially ab-based and essentially ab-monotonic if and only if it is compatible with  $\overset{\{a, b\}}\sqsupset$.
\end{prop}
\dimo
{\rm Let us assume the compatibility. In order to prove  that $\phi$ is essentially ab-based and ab-monotonic, we show the implication $(1')$:

$$
 \left[ 
\begin{array} {ll}
P, Q\in\cP; \, v\in I(P)\cap I(Q)\Rightarrow P_v=Q_v; \, \phi(P)=a; \\
D(a,Q)\supseteq D(a,P); 
D(b, P)\supseteq D(b, Q)  \\
\end{array}
\right]\Rightarrow  \phi(Q)=a.
 $$

Once we notice that we cannot find voters $v$ in $I(P)$   for which $b\underset {Q_v}\succ a$, whereas for agents $v\in I(P)$  if we have that $a\underset {Q_v}\sim b$,
then $P_v=Q_v$,  we can represent the left hand side of the implication we have to prove as in the figure below.

 {\small
$$
\begin{tabular}{llllllll}
\cline{2-7}
Profile $P$ & \multicolumn{1}{|l}{$D(a,P)$} & \multicolumn{2}{|l}{$\ \ \ \ \
\ \ I(P)$} & \multicolumn{3}{|l}{$\ \ \ \ \ \ \ \ \ \ D(b,P)$} & 
\multicolumn{1}{|l}{$\overset\phi\longrightarrow a$} \\ \cline{2-7}
&  &  &  &  &  &  &  \\ \cline{2-7}
Profile $Q$ & \multicolumn{1}{|l}{$a\succ b$} & \multicolumn{1}{|l}{$P=Q$} & 
\multicolumn{1}{|l}{$a\succ b$} & \multicolumn{1}{|l}{$a\succ b$} & 
\multicolumn{1}{|l}{$b\succ a$} & \multicolumn{1}{|l}{$a\sim b$} & 
\multicolumn{1}{|l}{} \\ \cline{2-7}
&  & \multicolumn{1}{|l}{} & \multicolumn{1}{|l}{} & \multicolumn{1}{|l}{} & 
\multicolumn{1}{|l}{} & \multicolumn{1}{|l}{} &  \\ 
&  & \multicolumn{1}{|l}{$\ \ \ \ V_{1}$} & \multicolumn{1}{|l}{$\ \ V_{2}$}
& \multicolumn{1}{|l}{$\ \ V_{3}$} & \multicolumn{1}{|l}{$\ \ \ V_{4}$} & 
\multicolumn{1}{|l}{$\ \ \ V_{5}$} &  \\ 
&  & \multicolumn{1}{|l}{} & \multicolumn{1}{|l}{} & \multicolumn{1}{|l}{} & 
\multicolumn{1}{|l}{} & \multicolumn{1}{|l}{} &  \\ \cline{2-7}
Profile $R$ & \multicolumn{1}{|l}{$a\succ b$} & \multicolumn{2}{|l}{$\ \ \ \
\ R=P$} & \multicolumn{1}{|l}{$a\sim b$} & \multicolumn{1}{|l}{$b\succ a$} & 
\multicolumn{1}{|l}{$R=Q$} & \multicolumn{1}{|l}{} \\ \cline{2-7}
\end{tabular}
$$}


Let us introduce a new profile $R$. With reference to the partition $\{ D(a, P), V_1, V_2, V_3, V_4, V_5 \}$ of the society $V$, we take any profile that ensures: on
$D(a, P)$ that $a\underset{R_v}\succ b$, on $I(P)=V_1\cup V_2$ is identical to $P$, on $V_3$ ensures that $a\underset{R_v}\sim b$, on $V_4$ ensures that $b\underset{R_v}\succ a$, is $Q$ on the set $V_5$.

It is straightforward to check that $$V=E(R,P)\cup [I(R)\cap D(b,P)]=E(R,Q)\cup [I(R)\cap D(a,Q)],$$
what means that the profile $P$ is $b$-dominated by $R$ and the profile $Q$ is $a$-dominated by $R$. By the compatibility assumption we first have that $\phi(R)=a$ and then that $\phi(Q)=a$ as desired. 

For the converse, we take into account that we can assume the validity of the implications $(1)$ and $(2)$ above, without requiring the strictness of the inclusions involved.  Let us suppose now that $P\underset{\phi(P)}\sqsupset Q$. We have to obtain that $\phi(Q)=\phi(P)$. Without loss of generality consider $\phi(P)=a.$ We observe that we must have $D(a,Q)\supseteq D(a, P)$. Indeed, if this is not the case we find, $v$ with $a\underset{P_v}\succ b$ and either $a\underset{Q_v}\sim b$ or $b\underset{Q_v}\succ a$. Since $V=E(P,Q)\cup [I(P)\cap D(a,Q)],$ and $v\notin E(P,Q)$ we get a contradiction.}}
\eproof
 
\end{oss}

\section{The  classes $\cP_a(\pi, \cF)$ and $\cP_b(\pi, \cF)$ of profiles, and the  index of a profile}
\lhead{\sc\scriptsize The structure of two-valued ...}
\rhead{\sc\scriptsize The classes $\cP_a(\pi, \cF)$, $\cP_b(\pi, \cF)$ and the index}

In \cite[Definition 2.6]{KPS2}, we defined superset closed families of coalitions of voters (also known as committees in the Social Choice literature) as nonempty subsets $\cF$ of $2^V\setminus\{\O\}$ that are closed under supersets, i.e. $E\supseteq F\in \cF\Rightarrow E\in \cF.$ It will be useful to have a slight extension of this concept for our case of  two-valued, rather than binary, functions.  

\begin{defin}
$\cF$ is a {\bf superset closed family on a set $V$} - and we write $\cF\in SSCF(V)$ - if: 
$\cF\subseteq 2^V$,  and $E\supseteq F\in \cF\Rightarrow E\in \cF.$ 

 We shall write $\cF\in SSCF$ to mean that for some subset $D$ of $V$, $\cF\in SSCF(D)$. 
\end{defin}

Notice that if $\cF$ is empty, then it is a superset closed family. 

According to the fact that an  $\cF\in SSCF(V)$ which is nonempty either contains $\O$ or not, we obviously have: 
in the first case $\cF$ is necessarily the power set $2^V$; 
in the second case it is a superset closed family of coalitions, the largest one being $2^V\setminus\{\O\}$.

If $\cF\in SSCF(V)$, setting, as in \cite{KPS2}, $E\in\cF^\circ\Leftrightarrow E^c\notin \cF$ we define {\bf the dual of $\cF$}. 

Trivially: $\cF^\circ\in SSCF(V)$;  $\cF^{\circ\circ}=\cF$; the empty family ($\cF=\O$) and the power set ($\cG=2^V$) are dual  to each other; the dual of a superset closed family of coalitions is also a superset closed family of coalitions.

 In other words to the dual pairs $\langle \cF, \cF^\circ\rangle$ of superset closed family of coalitions considered, solely, in \cite{KPS2} we have just added the pairs $\langle \O, 2^V\rangle$, $\langle 2^V, \O\rangle$. 

\bigskip
Fix two distinct alternatives $a, b\in A$.  We shall now define two useful classes of profiles, denoted by 
$\cP_a(\pi, \cF)$ and $\cP_b(\pi, \cF)$, for $\pi\in\cP(a\sim b)$ and $\cF\in SSCF(V\setminus {\rm dom}(\pi))$.

\begin{defin}\label{classe}
Let $\pi$ a partial $\{a, b\}$-indifference profile whose domain is the set $I$. Let $\cF$ a superset closed family on $I^c$.

We say that {\bf a profile  $P\in\cP$ belongs to the class $\cP_a(\pi, \cF)$ if} the following two conditions hold:

$\left\{ 
\begin{array} {ll}
D(a, P)\cap I^c \in {\cal F}; \\

v\in I\Rightarrow \mbox{ either }P_v=\pi_v, \mbox{ or  the restriction of  } P_v \mbox{ to } \{a, b\} \mbox{ is } a\succ b.
\end{array}
\right.$

\bigskip
Analogously, {\bf we say that $P\in\cP$ belongs to  $\cP_b(\pi, \cF)$ if} :

$\left\{ 
\begin{array} {ll}
D(b, P)\cap I^c \in {\cal F}^\circ; \\

v\in I\Rightarrow \mbox{ either }P_v=\pi_v, \mbox{ or  the restriction of  } P_v \mbox{ to } \{a, b\} \mbox{ is } b\succ a.
\end{array}
\right.$

\end{defin}

A pair $(\pi, \cF)$ can be thought of as an extension of the concept of superset closed family (hence, of committee). With this in mind, Definition \ref{classe} can be interpreted as an extension of preferences to
 $(\pi,\cF)$ if one says (see point 4. in the next Remark), given the profile $P$, that the extended committee $(\pi,\cF)$ prefers $a$ to $b$ if $P\in \cP_a(\pi, \cF)$, prefers $b$ to $a$ if $P\in \cP_b(\pi, \cF)$, and it is indifferent if $P\notin [\cP_a(\pi, \cF)\cup\cP_b(\pi, \cF)]$.

\begin{oss}\label{utile}
{\sl  Note that: 
\begin{enumerate}
\item[1.]   $\cP_a(\pi, \O)$ is empty;\quad
$P\in\cP_b(\pi, \O)$ iff $v\in I\Rightarrow \mbox{ either }P_v=\pi_v, \mbox{ or } b\underset {P_v} \succ a;$

\item[2.]  $P\in\cP_a(\pi, 2^{I^c})$ iff $v\in I\Rightarrow \mbox{ either }P_v=\pi_v, \mbox{ or } a\underset {P_v} \succ b;$ \quad $\cP_b(\pi, 2^{I^c})$ is empty.
\end{enumerate}
\noindent
The relations above cover the case that $\pi$ is a (total) profile of unanimous indifference between $a$ and $b$. Indeed in this case $I^c$ is empty and the possible choices for $\cF$ on $I^c$ are only two (each other dual): $\cF$ is empty,  or is the power set of $I^c$ (i.e. $\cF=\{\O\}$).
\begin{enumerate}
\bigskip
\item[3.]  If $\pi$ is the empty profile, and hence $\cF\in SSCF(V)$, we have:
$P\in\cP_a(\O, \cF)$ iff $D(a, P)\in \cF$; \quad $P\in\cP_b(\O, \cF)$ iff $D(b, P)\in \cF^\circ.$ 

\bigskip
\item[4.] The classes $\cP_a(\pi, \cF)$ and $\cP_b(\pi, \cF)$  are disjoint. Indeed, suppose $P\in \cP_a(\pi, \cF)\cap\cP_b(\pi, \cF)$. This gives  the validity of the conditions:
$D(a, P)\cap I^c\in\cF$; 
\,\,
$D(b, P)\cap I^c\in\cF^\circ$.
In particular: $D(b, P)\cap I^c\in\cF^\circ$ is the same as $I^c\setminus \big[D(b, P)\cap I^c\in\cF^\circ \big]\notin\cF$. But the latter set is $I^c\setminus D(b, P)$ and includes   $D(a, P)\cap I^c\in\cF$, a contradiction.

\medskip
\item[5.] In case $A$ contains only the two alternatives $a$ and $b$,   for the conditions considered in the Definition \ref{classe} above to define the classes $\cP_a(\pi, \cF)$ and $\cP_b(\pi, \cF)$, the second lines reduce simply to $D(b, P)\cap I=\O$ and to  $D(a, P)\cap I=\O$ respectively.\end{enumerate} }

\end{oss}

For an ordinal $\beta\geq 1$  (finite or not), let $\Lambda$ denote the well ordered set  $\{\lambda:  {0 \leq \lambda < \beta}\}$. For every $\lambda \in \Lambda$, let $\pi^{\lambda}$ denote a partial $\{a, b\}$-indifference profile.

We call {\bf well ordered collection of partial $\{a, b\}$-indifference profiles}
  a map    {\boldmath$\Pi$}$=(\pi^\lambda)_{\lambda\in \Lambda}=\{\pi^\lambda: 0\leq\lambda<\beta\}$    from a well ordered set to $\cP(a\sim b)$. Similarly, a map {\boldmath$\cF$} from a well ordered set to $SSCF$ will be called
{\bf well ordered collection of superset closed families}.

\begin{defin}\label{doublecollection}

Let $a$ and $b$ two distinct alternatives. We call {\bf double collection with respect to $\{a, b\}$}  a pair $\langle${\boldmath$\Pi$}, {\boldmath$\cF$}$\rangle$ consisting  of a 
well ordered collection {\boldmath$\Pi$} of partial $\{a, b\}$-indifference profiles, and 
a well ordered collection {\boldmath$\cF$} of superset closed families such that:
\begin{itemize}
\item [(i)] {\boldmath$\Pi$} and  {\boldmath$\cF$} have a common well ordered domain $\Lambda$, i.e.  
{\boldmath$\Pi$}$=(\pi^\lambda)_{\lambda\in \Lambda}$ and 
{\boldmath$\cF$}$=(\cF_\lambda)_{\lambda\in \Lambda}$

\item[(ii)] $\cF_\lambda\in SSCF(I^c_\lambda)$ for every $\lambda\in \Lambda$, where $I_\lambda=$ {\rm dom}$\,(\pi^\lambda)$.
 
\end{itemize}
\end{defin}
It is possible to describe a double collection also as a well ordered collection of extended committees.

We are now ready to introduce  the notion of index of a profile. This notion is more general than the corresponding notion considered in \cite{KPS2} for the case of $A = \{a,b\}$, and is the key tool to describe the sequential structure of CSP scfs..

\begin{defin}\label{indice}
Given the two distinct alternatives $a$ and $b$,
and  a double collection with respect to $\{a, b\}$ $\langle (\pi^\lambda)_{\lambda\in \Lambda}, (\cF_\lambda)_{\lambda\in \Lambda}\rangle$, {\bf the index of a profile $P$}, denoted by $\lambda(P)$, is:
 \begin{itemize}
 \item[ ]
the first element of the set
$
\{\lambda\in \Lambda: P\in \cP_a(\pi^\lambda, \cF_\lambda) \cup \cP_b(\pi^\lambda, \cF_\lambda)\},
$
 if it is nonempty; 

\item [ ]
$\infty$, if the above set is empty.\end{itemize}
\end{defin}


A symbol like $\lambda_{\{a, b\}}(P)$ for the index of $P$, and the name of $\{a, b\}$-index  would be more appropriate{\footnote{But we should take into consideration even the double collection.}}. However, we keep the shorter 
$\lambda(P)$,  since throughout the sequel there will be no ambiguity. Notice that if, as in \cite{KPS2}, there are only two alternatives $a$ and $b$, since in this case partial indifference profiles can be identified with their domain, the concept of double collection here, coincide with that in \cite{KPS2}. Moreover, by 5. in Remark \ref{utile}, Definition \ref{indice} above reduces to \cite[Definition 3.1]{KPS2}.

\section{Constructing strategy-proof social choice functions} 
\lhead{\sc\scriptsize The structure of two-valued ...}
\rhead{\sc\scriptsize Constructing CSP scfs}

In this section we introduce a formula (formula $(\bigstar)$ below) for the construction of CSP scfs whose ranges are of cardinality two.  This formula makes use  of the index $\lambda$ defined before. We shall see in the next section that every CSP scf with a range of cardinality at most two can be constructed in this way. In other words the formula we are speaking about will give a representation (next Theorem \ref{rappresentazione}) for  all two-valued CSP social choice functions. 

\bigskip
 \begin{defin}\label{defofpsi}
A scf will be said to be of $\psi${\bf -type}  if 
for a suitable choice of: 
\begin{itemize}
\item $\{a, b\}\subseteq A$,

\item  $x\in\{a, b\}$, and 
 
\item  a double collection $\langle${\boldmath$\Pi$}, {\boldmath$\cF$}$\rangle$ with respect to $\{a, b\}$,
 \end{itemize}
for every profile $P$ the corresponding value is determined as follows:
 
 $$
 (\bigstar) \qquad P\in \cP\mapsto 
 \left\{ 
\begin{array} {ll}
 a,   & \mbox{ if }   P\in {\cal P}_a ( \pi^{\lambda(P)}, \cF_{\lambda(P)}) \\
 b,   & \mbox{ if } P\in {\cal P}_b ( \pi^{\lambda(P)}, \cF_{\lambda(P)}).  \\
 x,   & \mbox{ if } \lambda(P)=\infty \\
\end{array}
\right.
 $$

A scf of $\psi$-type  will be denoted by $\psi$, omitting, for our sake of notational simplicity, the reference to  $\{a, b\}, \, x$ and $\double$.
\end{defin}

{\sl

{\bf The function $\phi$ of Example \ref{DIA}.}

Let us consider the function $\phi$ of Example \ref{DIA}. We recognize it as defined by means of $(\bigstar)$ if we take:
\begin{itemize}
\item the partial $\{a, b\}$-indifference profiles $\pi^0,\,\pi^1$ defined both on $\{v_1\}$ respectively as follows: 
$
 a \underset{\pi^0_{v_1}}\sim b \underset{\pi^0_{v_1}} \succ c; \qquad 
c \underset{\pi^1_{v_1}} \succ a \underset{\pi^1_{v_1}}\sim b;
$
\item the corresponding families $\cF_0$ and $\cF_1$ on $\{v_2\}$ respectively as follows:
$
\cF_0=\{\{v_2\}\}=\cF_0^\circ; \qquad \cF_1= {\rm empty, }\,\, \cF_1^\circ=\{\O,\{v_2\}\} 
$.
\end{itemize}
Then we see that 
\begin{itemize}
\item []a profile $P$ belongs to $\cP_a(\pi^0,\cF_0)$ iff $a\underset{P_{v_2}}\succ b$ and either one of the following is true:  $P_{v_1}=\pi^0_{v_1}$ or $a\underset{P_{v_1}}\succ b;$

\item []a profile $P$ belongs to $\cP_b(\pi^0,\cF_0)$ iff $b\underset{P_{v_2}}\succ a$ and either one of the following is true:  $P_{v_1}=\pi^0_{v_1}$ or $b\underset{P_{v_1}}\succ a;$

\item []$\cP_a(\pi^1,\cF_1)$ is empty whereas a profile $P$ belongs to $\cP_b(\pi^1,\cF_1)$ iff either one of the following is true:  $P_{v_1}=\pi^1_{v_1}$ or $b\underset{P_{v_1}}\succ a.$
\end{itemize}

Finally, the function $\phi$ is the $\psi$-type scf associated to the double collection
$\langle (\pi^0,\pi^1), (\cF_0,\cF_1)\rangle$ and to $x=a$.\eproof

\bigskip
}

 \begin{prop}\label{range}
Suppose $\psi$ is associated to $x\in\{a, b\}$, and $\double $. Suppose the double collection consists just of one element $\langle \pi^0, \cF_0\rangle$ (say $\beta$ is $1$). Then, the range of $\psi$:
\begin{itemize}
\item[(1)] contains $a$  and $x$, if $\cF_0$ is the power set of $I_0^c$ (and the dual family $\cF^\circ_0$ is empty)
\item[(2)] contains $b$  and $x$, if $\cF_0$ is empty   (and the dual family $\cF^\circ_0$ is the power set of $I^c_0$)
\item[(3)] is $\{a, b\}$, if both $\cF_0$ and $\cF^\circ_0$ are nonempty.
\end{itemize}
In cases $(1)$ and $(2)$ the scf $\psi$ may be constant.
\end{prop}
\dimo
In the argument we can omit the index 0 to $\pi^0$ and  $\cF_0$. Let $\pi$ be a partial $\{a, b\}$-indifference profile whose domain is the set $I$. Let $\cF$ be a superset closed family on $I^c$.
Notice that if $Q$ is a profile that ensures that $D(a, Q)=V$, then by Definition \ref{classe}
one has 
$$
\big[Q\in\cP_a(\pi,\cF)\Leftrightarrow I^c\in \cF\Leftrightarrow \cF \mbox{ nonempty }\big] \mbox{ and } Q\notin\cP_b(\pi,\cF).
$$
Similarly, if $R$ ensures that $D(b, R)=V$, then
$$
R\notin\cP_a(\pi,\cF) \mbox{ and }
\big[R\in\cP_b(\pi,\cF)\Leftrightarrow I^c\in \cF^\circ\Leftrightarrow \cF^\circ \mbox{ nonempty }\big].
$$
If the family $\cF$ is the power set of $I^c$, we saw (Remark \ref{utile}) that necessarily the class $\cP_b(\pi, \cF)$ is empty\footnote{Also: $P\in \cP_a(\pi, \cF)$ is the same as $\big[ v\in I\Rightarrow {\rm  either } \,\, P_v=\pi_v\,\, {\rm or } \,\,a\underset{P_v}\succ b  \big]$; notice that $\phi(\pi)=\phi(Q)=a$ and $\phi(P)=x$ if the profile $P\in\cP(a\sim b)$ and $P_I\neq \pi$.}, and this means that we cannot guarantee a priori  the value $b$ for $\psi$.
If the family $\cF^\circ$ is the power set of $I^c$, we saw that necessarily the class $\cP_a(\pi, \cF)$ is empty\footnote{Also: $P\in \cP_b(\pi, \cF)$ is the same as $\big[ v\in I\Rightarrow {\rm  either } \,\, P_v=\pi_v\,\, {\rm or } \,\,b\underset{P_v}\succ a  \big]$.}, and this means that we cannot guarantee a priori  the value $a$ for $\psi$.
Combining all the above remarks, the statements {\it (1), (2),} and {\it (3)} follow.
\eproof

\begin{corol}\label{duevalori}
Suppose $\psi$ is associated to $x\in\{a, b\}$, and $\double $. Suppose that  both $\cF_{0}$ and $\cF^\circ_{0}$ are nonempty, then, the range of $\psi$ is $\{a, b\}$.
\end{corol}

Notice that in \cite{KPS2} it is always the case that for the first index
$0$ of $\Lambda$  both $\cF_{0}$ and $\cF^\circ_{0}$ are nonempty. The $\psi$-type functions introduced here specialize to those defined in \cite{KPS2}.

We now have the main result of this section.

\begin{theo}\label{psiiscsp}
Every $\psi$-type function is coalitionally strategy-proof.
\end{theo}

\dimo
Suppose $\psi$ is associated to $x\in\{a, b\}$, and the double collection $\double$ with respect to  
$\{a, b\}$. We assume the range is $\{a, b\}$. By appealing to Theorem \ref{final2} we have to show that
\quad
$
P\underset {\psi(P)}\sqsupset Q\,\Rightarrow \,\, \psi(Q)=\psi(P). 
$

\bigskip
With this purpose in mind,  we  prove the following two claims under the assumption that  $P\underset b\sqsupset Q$,

\begin{itemize}
\item [] (claim 1) $\qquad \big[\lambda(P)<\infty, \, \psi(P)=b \big]\,\Rightarrow \big[Q\in \cP_b(\pi^{\lambda(P)}, \cF_{\lambda(P)})\big]$.

\item[] (claim 2)  $\qquad \big[\lambda(Q)<\infty, \, \psi(Q)=a \big]\,\Rightarrow \big[P\in \cP_a(\pi^{\lambda(Q)}, \cF_{\lambda(Q)})\big]$.
\end{itemize}

The assumption that $P\underset b\sqsupset Q$, i.e. $V=E(P,Q)\cup\big[I(P)\cap D(b, Q)\big]$, means that for every voter $v$ one of the following four circumstances holds true:

(i) $P_v=Q_v$;

(ii) the restrictions to $ \{a, b\} $ of the preferences $P_v$ and $Q_v$ are identical and coincide with $a\succ b$;

(iii) the restrictions to $ \{a, b\} $ of the preferences $P_v$ and $Q_v$ are identical and coincide with $b\succ a$;

(iv) \quad $a\underset{P_v}\sim b$ and \,\, $ b\underset{Q_v}\succ a$.

{\sc proof of claim 1:}
By definition of $\psi$, we have that $P\in \cP_b(\pi^{\lambda(P)},\cF_{\lambda(P)})$. In its turn this means:

$$\left\{ 
\begin{array} {ll}
D(b, P)\cap I^c_{\lambda(P)} \in {\cal F}^\circ_{\lambda(P)} \\

[*]\,\,v\in I_{\lambda(P)}\Rightarrow \mbox{ either }P_v=\pi^{\lambda(P)}_v, \mbox{ or  the restriction of  } P_v \mbox{ to } \{a, b\} \mbox{ is } b\succ a.
\end{array}
\right.$$
Since for voters in $D(b, P)$ we can only have either (i) or (iii), we recognize that $D(b, P)\subseteq D(b, Q)$, and by the superset closedness  we have
$$
D(b, Q)\cap I^c_{\lambda(P)} \in {\cal F}^\circ_{\lambda(P)}.
$$
We also have that
$$[**]\,\, v\in I_{\lambda(P)}\Rightarrow \mbox{ either }Q_v=\pi^{\lambda(P)}_v, \mbox{ or  the restriction of  } Q_v \mbox{ to } \{a, b\} \mbox{ is } b\succ a,
$$
and these two facts give  the assertion of claim 1. 

To check $[**]$, notice that if $v\in I_{\lambda(P)}$ and $P_v\neq Q_v$, because of $[*]$ either (iii) or (iv) is true and consequently the restriction of  $Q_v \mbox{ to } \{a, b\} \mbox{ is } b\succ a.$\eproof

\bigskip
{\sc proof of claim 2:}  
Quite similarly to what before, by definition of $\psi$, we have that $Q\in \cP_a(\pi^{\lambda(Q)},\cF_{\lambda(Q)})$. In its turn this means:

$$\left\{ 
\begin{array} {ll}
D(a, Q)\cap I^c_{\lambda(Q)} \in {\cal F}_{\lambda(Q)} \\

[*']\,\,v\in I_{\lambda(Q)}\Rightarrow \mbox{ either }Q_v=\pi^{\lambda(Q)}_v, \mbox{ or  the restriction of  } Q_v \mbox{ to } \{a, b\} \mbox{ is } a\succ b.
\end{array}
\right.$$
Since for voters in $D(a, Q)$ we can only have either (i) or (ii), we recognize that $D(a, Q)\subseteq D(a, P)$, and by the superset closedness  we have
$$
D(a, P)\cap I^c_{\lambda(Q)} \in {\cal F}_{\lambda(Q)}.
$$
We also have that
$$[**']\,\, v\in I_{\lambda(Q)}\Rightarrow \mbox{ either }P_v=\pi^{\lambda(Q)}_v, \mbox{ or  the restriction of  } P_v \mbox{ to } \{a, b\} \mbox{ is } a\succ b,
$$
and these two facts give  the assertion of claim 2.

For $[**']$, notice that if $v\in I_{\lambda(Q)}$ and $P_v\neq Q_v$, because of $[*']$ we have  (ii)  and consequently the restriction of  $P_v \mbox{ to } \{a, b\} \mbox{ is } a\succ b.$\eproof

We can now  move to the proof of 
$$
P\underset {\psi(P)}\sqsupset Q\,\Rightarrow \,\, \psi(Q)=\psi(P), 
$$
in the case that $\psi(P)=b$. We are not  losing any generality in this since the case that $\psi(P)=a$ can be treated similarly by means of an obvious modification of claims 1 and 2.

Suppose the index $\lambda(P)$ is finite. Then we have by claim 1 that $\lambda(Q)\leq \lambda(P)$. Consequently if $\psi(Q) =a$, by claim 2 we get  $\lambda(Q)= \lambda(P)$ and also
$P\in \cP_a(\pi^{\lambda(P)}, \cF_{\lambda(P)})$ which contradicts the assumption that $\psi(P)=b$.

Suppose now that $\lambda(P)=\infty$. This means, by definition of $\psi$, that $\psi(P)=b=x$. It follows that if $\psi(Q)=a$,  it must be true that $\lambda(Q)<\infty$, then again claim 2 applies and $\lambda(P)$ should be finite. 
\eproof


\section{The structure theorem} 
\lhead{\sc\scriptsize The structure of two-valued ...}
\rhead{\sc\scriptsize Structure theorem}

In this section we shall see that every  CSP scf $\phi$ defined on all weak profiles, 
whose range is $\{a, b\}$, can be represented by means of the formula $(\bigstar)$ in Definition \ref{defofpsi}. In order to achieve this result 
we shall first identify  particular classes of profiles based on top choice  only (choice between $a$ and $b$), that necessarily have to go to prescribed values. This is essentially the content of the following lemma.

Let $\{\phi=x\}$, with $x\in A$, denote the set of all profiles $P$ attaining value $x$ under $\phi$, i.e. $\phi(P)=x$.

\begin{prop}\label{truetrue}
Let $\phi:\cP\to A$ be  CSP. Assume that the   range of $\phi$ is $\{a, b\}$. Then there exists a unique
 superset closed family $\cal F$ of coalitions such that, if $\pi$ is the empty profile,  one has:\quad 
 $\cP_a(\pi, \cF)\subseteq \{\phi=a\}$, \, and \quad
 $\cP_b(\pi, \cF)\subseteq \{\phi=b\}$.




\end{prop}

\dimo
Take the family $\cF$ accordingly to  Theorem \ref{propo16} $(i)$ applied to  the restriction of $\phi$ to the subset of $\cP$ of all strict profiles.
We show the  implication $$ (+) \qquad D(a, P)\in {\cal F}\Rightarrow \phi(P)= a.$$ Similarly one can obtain  the other implication 
$$ (++) \qquad D(b, P)\in {\cal F}^\circ\Rightarrow \phi(P)= b.$$ 

Let $P$ be an arbirary profile. 
We can assume that the coalition $D(a, P)$ is a proper subset of $V$ \footnote{Otherwise we just use the Pareto optimality (see Remark \ref{Pareto}).}. Let us take two strict orderings ${S_1}$ and ${S_2}$ on $A$ that guarantee $a\underset{{S_1}}\succ b$ \, and \, 
$b\underset{{S_2}}\succ a$, and define a strict profile $\widehat P$ in the following way.

$$\widehat P_v=\left\{ 
\begin{array} {ll}
S_1, \mbox{ for } v\in D(a, P) \\

S_2, \mbox{ for } v\notin D(a, P).
\end{array}
\right.$$
Since $D(a, P)=D(a, \widehat P)$ and the implication we need to prove  is true for strict profiles (as Theorem \ref{propo16} $(i)$ says), we have that $\phi(\widehat P)=a$. We also have  $\phi([P_{D(a,P)}, \widehat P_{D(a,P)^c}])=a$, otherwise the coalition $D(a,P)$ manipulates the profile 
$[P_{D(a,P)}, \widehat P_{D(a,P)^c}]$ by presenting $\widehat P$. We can conclude 
$\phi(P)=a$ as desired since otherwise the coalition $D(a,P)^c$ manipulates the profile 
$[P_{D(a,P)}, \widehat P_{D(a,P)^c}]$ by presenting $P$.

According to point 3. of Remark \ref{utile}, we can rewrite the above implications $(+)$ and $(++)$  respectively as $\cP_a(\pi, \cF)\subseteq \{\phi=a\}$, \, and \quad
 $\cP_b(\pi, \cF)\subseteq \{\phi=b\}$. For the uniqueness of $\cF$, the same argument used in the proof of \cite[Proposition 2.9 ]{KPS2} \footnote{For the uniqueness of the family there called $\cF_0.$}, applies.
\eproof


We are now ready to state our representation theorem. It says that all  CSP scfs  are necessarily scfs of $\psi$-type. 

Let us recall that: a profile of unanimous indifference between $a$ and $b$ is a profile  under which all voters are indifferent between the two alternatives.

\begin{theo}\label{rappresentazione}
Let $\phi:\cP\to A$ be a scf which is coalitionally strategy-proof and has range $\{a, b\}\subseteq A$ of cardinality two. Let $\pi$ be a profile of unanimous indifference between the alternatives $a$ and $b$, and set $x=\phi(\pi)$. 

Then, we can find a double collection $\double=\langle (\pi^\lambda)_{0\leq\lambda<\beta},  (\cF_\lambda)_{0\leq\lambda<\beta} \rangle$ with respect to $\{a, b\}$  such that the function $\phi$ coincides with the $\psi$-type function associated  with $\double$ and  $x$. 

Moreover, the dual families
$\cF_{0}$ and $\cF_{0}^\circ$ are both nonempty, {\footnote {Namely,  they consist of coalitions. \\ Naturally, compare Theorem \ref{rappresentazione} with the combination of Corollary \ref{duevalori} and Theorem \ref{psiiscsp}}
} and the ordinal  $\beta$ is finite if $V$  and $A$ are both finite.
\end{theo}

\section{Appendix: the proof of Teorem \ref{rappresentazione}}
\lhead{\sc\scriptsize The structure of two-valued ...}
\rhead{\sc\scriptsize Appendix}
The proof  will be achieved by transfinite induction. Before going into the details of the proof, it is convenient to adopt some  notation.

\begin{itemize}
\item[$(N_1)$] The elements of $\cP$, i.e. all profiles,  will be enumerated as $P^0, P^1, \cdots, P^{\eta}, \cdots : \eta < \delta$ where $\delta$ is some ordinal.  We shall denote  by $\cal E$ this enumeration. In other words we have $\cP=\{P^\eta: \eta<\delta\}$ for a suitable ordinal $\delta$. 
Note that $\delta$ is finite if $V$ and $A$ are finite. To fix ideas we can assume that $P^0$ is the profile $\pi$.
\item[$(N_2)$] If we have a partial $\{a, b\}$-indifference profile $\pi^\lambda$, and $\cF_\lambda\in SSCF(V\setminus$ dom$(\pi^\lambda))$, for the set of profiles $\cP_a(\pi^\lambda, \cF_\lambda) \cup \cP_b(\pi^\lambda, \cF_\lambda)$ we shall use the notation $\cP_\lambda$, for the sake of simplicity.
\item[$(N_3)$] For every ordinal $\alpha$, the set $\Delta_\alpha$ is defined as
$$
\big\{ \eta<\delta: P^\eta\in \cP\setminus \big(\bigcup_{0\leq\lambda<\alpha}\cP_\lambda\big)
\mbox{ and } \phi(P^\eta)\neq x\big\}=:\Delta_\alpha.
$$
\item[$(N_4)$] When $\Delta_\alpha\neq\O$, $Q^\alpha$ stands for the profile whose index in the enumeration $\cal E$ is the first element of $\Delta_\alpha$.  In other words $Q^\alpha=P^{\eta_\alpha}$, where $\eta_\alpha:=$ min $\Delta_\alpha$. 
\end{itemize}

Notice that if $\alpha<\beta$, then $\Delta_\alpha\supseteq\Delta_\beta$, hence $\eta_\alpha\leq\eta_\beta$.

\bigskip

The following is the key technical tool in order to obtain our representation  theorem stated in Theorem \ref{rappresentazione}. 

\begin{lem}\label{mainlemma}
Let $\phi:\cP\to A$ be a scf which is coalitionally strategy-proof and has range $\{a, b\}\subseteq A$ of cardinality two. Let $\pi$ be a profile of unanimous indifference between the alternatives $a$ and $b$, and set $x=\phi(\pi)$. Assume that 

\begin{itemize}

\item [(i)] $\pi^0$ is the empty profile, $\cF_0$ is a superset closed family of coalitions on $V$ with the following property:
$$\Pa\supseteq \cP_a(\pi^0, \cF_0)\qquad \mbox{ and } \qquad
\Pb\supseteq \cP_b(\pi^0, \cF_0)$$

\item [(ii)] $\beta$ is an ordinal,
\item[(iii)]
$0<\alpha<\beta\Rightarrow \Delta_\alpha\neq\O$, and $\alpha\leq\eta_\alpha$,

\item [(iv)] for every $0<\alpha<\beta$, correspondingly to the profile $Q^\alpha$, there exists a  family $\cF_\alpha\in SSCF(V\setminus I(Q^\alpha))$ such that, setting
$I_\alpha:=I(Q^\alpha), \,\, \pi^\alpha_v:=Q^\alpha_v\,\, \mbox{ (for all $v\in I_\alpha$),  } $
we have:

$$ Q^\alpha\in\cP_\alpha\footnote{Hence, by the definition $(N_4)$ of the profile $Q^\alpha$, we have $Q^\alpha\in\cP_\alpha\setminus \big(\bigcup_{0\leq\lambda<\alpha}\cP_\lambda\big)$.}, \,\, \pi\notin \bigcup_{0\leq\alpha<\beta}\cP_\alpha, $$ $$
\Pa\supseteq \bigcup_{0\leq\alpha<\beta}\cP_a(\pi^\alpha, \cF_\alpha),
\, \mbox{ and }\,\,
\Pb\supseteq \bigcup_{0\leq\alpha<\beta}\cP_b(\pi^\alpha, \cF_\alpha).
$$
\end{itemize}
Under the above assumptions we have that either one of the following holds true:
\begin{itemize}
\item [1.] $\Delta_\beta=\O$, and the function $\phi$ coincides with the $\psi$-type function associated with $\double$ and  $x$, 
where the double collection $\double$ consists of the $\pi^\alpha$, and the $\cF_\alpha$ for all
$0\leq\alpha<\beta$;  
\item [2.] $\Delta_\beta\neq\O$, and for $Q^\beta$ there exists a  family $\cF_\beta\in SSCF(V\setminus I(Q^\beta))$ such that, setting
$I_\beta:=I(Q^\beta), \quad \pi^\beta_v:=Q^\beta_v\,\, \mbox{ (for all $v\in I_\beta$)  }, 
$
we have:

$$Q^\beta\in \cP_\beta\setminus \big(\bigcup_{0\leq\lambda<\beta}\cP_\lambda\big), \qquad Q^\beta\notin \{Q^\alpha: 0<\alpha<\beta\},\qquad 
\pi\notin\cP_\beta, \quad \beta\leq \eta_\beta,$$
$$\Pa\supseteq \cP_a(\pi^\beta, \cF_\beta),
\,\, \mbox{ and }\,\,
\Pb\supseteq \cP_b(\pi^\beta, \cF_\beta).
$$
\end{itemize}
\end{lem}

{\sc proof of Lemma \ref{mainlemma}}:
 We can distinguish two cases that will give rise to either 1. or 2. 

Suppose that  over $\cP\setminus(\bigcup_{0\leq\alpha<\beta} \cP_\alpha)$ the function $\phi$ takes one value only. This value is necessarily  $x$. Then what stated in 1. is obviously true, given the assumptions.

\bigskip
Hence, let us assume that over $\cP\setminus(\bigcup_{0\leq\alpha<\beta} \cP_\alpha)$ the function $\phi$ has range $\{a,b\}$. According to $(N_3)$ this is the same as $\Delta_\beta\neq\O$.

By definition $(N_4)$,
 $Q^\beta\in \cP\setminus(\bigcup_{0\leq\alpha<\beta} \cP_\alpha)$ and $\phi(Q^\beta)\neq \phi(\pi)$.  To fix ideas, we can assume, without loss of generality, that   $\phi(Q^\beta)=a.$ 

Set $I_\beta:=I(Q^\beta)$ and notice that this is a nonempty set \footnote{For a profile $P$ the condition $I(P)=\O$ gives $P\in\cP_0$. This is a consequence of  $(i)$ and what observed in point 3. of Remark \ref{utile}. }. Also notice that $Q^\beta\notin\{Q^\alpha: 0<\alpha<\beta\}$ since every $Q^\alpha \in \cP_\alpha$.
Since we know that $\eta_\alpha\leq\eta_\beta$, we entail that $\eta_\alpha<\eta_\beta$, otherwise we get $Q^\alpha=Q^\beta$ which is false.
Moreover $\beta\leq\eta_\beta.$ Indeed, if not, we have $ \beta>\eta_\beta$, then, setting $\eta_\beta=\alpha$, by assumption, from $\alpha<\beta$, we have $\alpha\leq\eta_\alpha$, hence $\eta_\beta\leq\eta_\alpha$, whereas we have just seen that the converse is true.

\bigskip
The restriction of $Q^\beta$ to $I_\beta$ is a partial $\{a,b\}$-indifference profile that we denote by $\pi^\beta$, so $\pi^\beta=(Q^\beta_v)_{v\in I_\beta}$.

We now associate to $\pi^\beta$ a family $\cF_\beta$  on $I_\beta^c$ according to the  following procedure. In case $V=I_\beta$ we set $\cF_\beta=\{\O\}$ (hence the dual family $\cF_\beta^\circ$ is empty). 
In case $I_\beta$ is properly contained in $V$, we distinguish two cases according to the cardinality of the range of the restriction 
$$ \qquad \phi_{I^c_\beta} (P_{I^c_\beta}):= \phi(\pi^\beta, P_{I^c_\beta})$$

of $\phi$ to the coalition $I_\beta^c$. 

Such a restriction alway takes value $a$ since $\phi_{I^c_\beta} (Q^\beta_{I^c_\beta})= \phi(\pi^\beta, Q^\beta_{I^c_\beta})=\phi(Q^\beta)$.
It may be possible that $b$ is never attained by means of the restriction. In this case we define $\cF_\beta$ as the power set of $I_\beta^c$ (hence the dual family $\cF_\beta^\circ$ is empty).

Notice that (see Remark \ref{utile}) in both cases considered so far, we have
$$P\in \cP_a(\pi^\beta, \cF_\beta) \mbox{ iff  for every voter $v\in I_\beta$ either }  P_v=\pi^\beta_v=Q^\beta_v \mbox{ or } a\underset{P_v}\succ b,$$
(whereas $\cP_b(\pi^\beta, \cF_\beta) \mbox{ is empty) }$
from which we get $$(\diamond)\qquad \cP_a(\pi^\beta, \cF_\beta)\subseteq \Pa.$$
Indeed, for every $P\in \cP_a(\pi^\beta, \cF_\beta)$ we have

\bigskip

{\footnotesize
$$\begin{tabular}{llll}
\cline{2-4}
Profile $P$ & \multicolumn{1}{|l}{$P$} & \multicolumn{1}{|l}{$Q^{\beta }$} & 
\multicolumn{1}{|l|}{$a\succ b$} \\ \cline{2-4}
& $I_{\beta }^{c}$ & $I_{\beta }^{\prime }$ & $I_{\beta }^{\prime
\prime }$%
\end{tabular}
$$}

where $I_\beta'=\{v\in I_\beta: P_v=Q^\beta_v\}$ and  $I_\beta''=\{v\in I_\beta: v\notin I_\beta' \,\,\&\,\, a\underset{P_v}\succ b\}$, and by strategy-proofness $\phi(P)=a$ otherwise the coalition $I_\beta''$ (only the case this set is nonempty is significant)  manipulates $P$ by presenting
$Q^\beta_{I_\beta''}.$

\bigskip
In case the function $\phi_{I^c_\beta}$ takes both values $a$ and $b$ we can apply Proposition \ref{truetrue} to this function determining a superset closed family $\cF_\beta$ of subcoalitions of $I_\beta^c$ such that

$$ D(a, P_{I_\beta^c})\in {\cal F}_\beta\Rightarrow \phi_{I_\beta^c}(P_{I^c_\beta})= a,$$

$$ D(b, P_{I_\beta^c})\in {\cal F}^\circ_\beta\Rightarrow \phi_{I_\beta^c}(P_{I_\beta^c})= b.$$
From Definition \ref{classe} we have again as before the inclusion $(\diamond)$ to which we can add the parallel inclusion
$$(\diamond')\qquad \cP_b(\pi^\beta, \cF_\beta)\subseteq \Pb.$$
Indeed, for every $P\in \cP_b(\pi^\beta, \cF_\beta)$ we have
{\footnotesize

\bigskip

$$\begin{tabular}{llll}
\cline{2-4}
Profile $P$ & \multicolumn{1}{|l}{$P$} & \multicolumn{1}{|l}{$Q^{\beta }$} & 
\multicolumn{1}{|l|}{$b\succ a$} \\ \cline{2-4}
& $I_{\beta }^{c}$ & $I_{\beta }^{\prime }$ & $I_{\beta }^{\prime
\prime }$%
\end{tabular}$$
}

where the only change is  that   $I_\beta''=\{v\in I_\beta: v\notin I_\beta' \,\,\&\,\, b\underset{P_v}\succ a\}$, and by strategy-proofness $\phi(P)=b$ otherwise the coalition $I_\beta''$ (only the case this set is nonempty is significant)  manipulates $P$ by presenting
$Q^\beta_{I_\beta''}.$

Having, at this stage, defined $\pi^\beta$ and $\cF_\beta$, 
by construction we have\footnote{See Definition \ref{classe}. Precisely, due to the assumption that $\phi(Q^\beta)=a$, the profile $Q^\beta\in  \cP_a(\pi^\beta,\cF_\beta)$.} \,\, $Q^\beta\in\cP_\beta=\cP_a(\pi^\beta,\cF_\beta)\cup \cP_b(\pi^\beta,\cF_\beta)$ and we have seen that $\cP_a(\pi^\beta,\cF_\beta)\subseteq \Pa$, and 
$\cP_b(\pi^\beta,\cF_\beta)\subseteq \Pb$.

To see that $\pi\notin\cP_\beta$, we notice what follows.
 
In case that $\cF_\beta$ is the power set of $I_\beta^c$:
then, because of formula ($\diamond)$ we have $\pi\notin \cP_a(\pi^\beta, \cF_\beta)$ and, on the other hand,   $\cP_b(\pi^\beta, \cF_\beta)$ is empty.

In case that $\cF_\beta$ comes from the application of Proposition \ref{truetrue} and, therefore, is a superset closed family of subcoalitions of $I^c_\beta$: observe that $D(a, \pi) \cap {I^c_\beta}=D(b, \pi) \cap {I^c_\beta}=\O$, so $\pi\notin \cP_a(\pi^\beta, \cF_\beta)$ and $\pi\notin \cP_b(\pi^\beta, \cF_\beta)$.
\eproof

We can now move to the proof of Theorem \ref{rappresentazione}.

\bigskip

{\sc proof of Theorem \ref{rappresentazione}:}
Setting $\lambda=0$, $I_0=\O$, and $\pi^0=$ {\it the empty profile}, we define the family $\cF_0$ by appealing to 
Proposition \ref{truetrue}. 

Notice that none of the profiles unanimously indifferent between $a$ and $b$ belongs to $\cP_0$.

We can now distinguish two cases according to the fact that:

\begin{itemize}

\item  Case 1: over $\cP\setminus \cP_0$ the function $\phi$ takes one value only.

\item Case 2: over $\cP\setminus \cP_0$ the function $\phi$ has range $\{a,b\}$.

\end{itemize}

Suppose we are in case 1. Then, the value attained by $\phi$ is necessarily $x$. Hence, the theorem is proved since $\phi$ coincides with the $\psi$-type function associated with 
the double collection $\langle \pi^0,  \cF_{0} \rangle$ and  $x$ (i.e. $\beta$ is one).

\bigskip
On the contrary, suppose that over $\cP\setminus \cP_0$ the function $\phi$ has range $\{a,b\}$. In this case if we set $\beta=2$, all the assumptions of Lemma \ref{mainlemma}
are satisfied. Indeed, note that the present case exactly says that $\Delta_1\neq\O$, and  what is needed to show is that for the profile $Q^1$ \footnote{Trivially $1\leq\eta_1$.} there is  a family $\cF_1\in SSCF(I_1^c)$ such that setting $\pi^1=Q^1_{I_1}$, we have $Q^1\in \cP_1$, $\pi\notin \cP_1$, 
$\Pa\supseteq \cP_a(\pi^1, \cF_1)$, and $\Pb\supseteq \cP_b(\pi^1, \cF_1)$.

\bigskip
The argument  is pretty much the same as in the proof of Lemma \ref{mainlemma}, and we summarize it in the following.

The set $I_1:=I(Q^1)$  is  nonempty. 
To the partial $\{a,b\}$-indifference profile  $\pi^1=(Q^1_v)_{v\in I_1}$
we  associate  a family $\cF_1$  on $I_1^c$ according to the  following procedure. In case $V=I_1$ we set $\cF_1=\{\O\}$ (hence the dual family $\cF_1^\circ$ is empty). 
In case $I_1$ is properly contained in $V$, we distinguish two cases according to the cardinality of the range of the restriction 
$$ \qquad \phi_{I^c_1} (P_{I^c_1}):= \phi([\pi^1, P_{I^c_1}])$$

of $\phi$ to the coalition $I_1^c$. 

Such a restriction always  takes value $\phi(Q^1)$ since $\phi_{I^c_1} (Q^1_{I^c_1})= \phi([\pi^1, Q^1_{I^c_1}])=\phi(Q^1)$.
It is possible that $x$ is never attained by $\phi_{I^c_1}$. In this case we define $\cF_1$ as the power set of $I_1^c$ (hence the dual family $\cF_1^\circ$ is empty).





\bigskip
In case the function $\phi_{I^c_1}$ takes both values $a$ and $b$ we can apply Proposition \ref{truetrue} to this function determining a superset closed family $\cF_1$ of subcoalitions of $I_1^c$ such that

$ D(a, P_{I_1^c})\in {\cal F}_1\Rightarrow \phi_{I_1^c}(P_{I^c_1})= a$, \quad and\quad
$ D(b, P_{I_1^c})\in {\cal F}^\circ_1\Rightarrow \phi_{I_1^c}(P_{I_1^c})= b.$





Having, at this stage,  defined $\cF_1$, we can verify that
 $$(\diamond)\qquad \cP_a(\pi^1, \cF_1)\subseteq \Pa,$$
$$(\diamond')\qquad \cP_b(\pi^1, \cF_1)\subseteq \Pb,$$
 $$Q^1\in\cP_1=\cP_a(\pi^1,\cF_1)\cup \cP_b(\pi^1,\cF_1), \mbox{ and } \pi\notin\cP_1.$$

 
%

\bigskip

Once the assumptions of Lemma \ref{mainlemma} have been proved, by this lemma, either we get that $\phi$ is the $\psi$-type function associated to the double collection $\langle (\pi^0,\pi^1), (\cF_0,\cF_1)\rangle$ and $x$, or it will be possible to repeat the application of Lemma \ref{mainlemma} (with $\beta=3$) since we shall have 
$\Delta_2\neq\O$ and the family $\cF_2$, the profile $\pi^2=Q^2_{I_2}$ with:

$Q^2\notin \{Q^1\}$,\, $2\leq \eta_2$,\,$Q^2\in\cP_2\setminus(\cP_0\cup\cP_1)$, \,$\pi\notin\cP_2$, \,
$ \cP_a(\pi^2, \cF_2)\subseteq \Pa, \,\, \cP_b(\pi^2, \cF_2)\subseteq \Pb.$

\bigskip
Then, either we  get that $\phi$ is the $\psi$-type function associated to the double collection $\langle (\pi^0,\pi^1, \pi^2), (\cF_0,\cF_1,\cF_2)\rangle$ and $x$, or we again can apply Lemma \ref{mainlemma} ($\beta=4$)\, ... We can  proceed by transfinite induction. 
The induction procedure will stop at an ordinal  $\gamma\leq\delta$ when the function $\phi$ takes only one value on the set $\big\{ \eta<\delta: P^\eta\in \cP\setminus \big(\bigcup_{0\leq\lambda<\gamma}\cP_\lambda\big)\big\}$. 

If $A$ and $V$ are finite the process clearly stops after finitely many steps.\eproof

\end{document}